\documentclass[numbered]{trbunofficial}
\usepackage{amsmath}
\usepackage{amssymb}
\usepackage{graphicx} 
\usepackage{mathrsfs}
\usepackage{tikz}
\usepackage{float}
\usepackage[colorlinks=false,linkcolor=blue,citecolor=blue]{hyperref}
\usepackage{subcaption}
\usetikzlibrary{shapes,arrows}

\graphicspath{ {./} }
\usepackage{subcaption}

\AuthorHeaders{Shuman, Guo and Caros}
\title{Journey-Based Transit Equity Analysis: A Case Study in the Greater Boston Area}

\author{\textbf{Daniela Shuman}\\
  School of Engineering and Applied Sciences \\
  Harvard University, Cambridge, MA 02138, USA \\
  Email: dshuman@college.harvard.edu\\
  \hfill\break
  \textbf{Xiaotong Guo}\\
  Department of Civil and Environmental Engineering \\
  Massachusetts Institute of Technology, Cambridge, MA 02139, USA \\
  Email: xtguo@mit.edu\\
  \hfill\break
  \textbf{Nicholas S. Caros}\\
  Department of Civil and Environmental Engineering \\
  Massachusetts Institute of Technology, Cambridge, MA 02139, USA \\
  Email: caros@mit.edu\\
  \hfill\break
}

\TotalWords{4369}

\begin{document}
\maketitle

\section{Abstract}

In this paper, a new methodology, journey-based equity analysis, is presented for measuring the equity of transit convenience between income groups.
Two data sources are combined in the proposed transit equity analysis: on-board ridership surveys and passenger origin-destination data. The spatial unit of our proposed transit equity analysis is census blocks, which are relatively stable over time and allows an exploration of the data that is granular enough to make conclusions about the service convenience various communities are facing.
A case study in the Greater Boston area using real data from the Massachusetts Bay Transportation Authority (MBTA) bus network demonstrates a significant difference in transit service convenience, measured by number of transfers per unit distance, transfer wait time and travel time per unit distance, between low-income riders and high income riders.
Implications of analysis results to transit agencies are also discussed.

\hfill\break
\noindent\textit{Keywords}: Public Transport, Origin-Destination Data, Spatial Equity Analysis
\newpage

\section{Introduction}


In urban areas across the globe, public transit systems provide mobility and access to opportunities that might not be available otherwise.
The mobility afforded by public transit is not always provided equitably, however.
Low-income residents generally experience worse access to public transit in major U.S. cities \cite{karner2015comparison, griffin2016public}.
In Amsterdam, low-income riders are more likely to have circuitous trips \cite{dixit2021examining}. 
Additionally, service cuts related to the COVID-19 pandemic disproportionately affected low-income communities in certain U.S. and Canadian cities \cite{deweese2020tale}. 

One of the key challenges in the transit equity analysis is the wide range of methods for defining and measuring equity.
In general, transit equity refers to an equitable distribution of transit services among communities and socio-economic groups. 
There are many different ways to quantify the quality of transit services, however, depending on the availability of data. 
In this study, demonstrate how rich transit origin-destination data at the individual trip level can be leveraged in order to measure the relative convenience of travel.
The chosen convenience measures are representative of two key dimensions of transit service quality that affect the passenger experience: in-vehicle time over distance, the number of transfers and transfer waiting time. 
The convenience metrics are compared between low- and high-income communities to determine whether transit service is equitable across income groups. 

Understanding the differences in transit service convenience between low and high-income areas is the first step towards achieving transit equity across the region.
First, these measures can serve as benchmarks and used to evaluate the impact of service changes with regards to equity. 
Additionally, the disaggregate nature of our proposed methods allows agencies to identify trips and routes that contribute to high in-vehicle time and transfer waiting time for low-income riders, thus enabling the prioritization of interventions to improve service quality.
For example, bus routes with high in-vehicle time could be addressed by adding transit priority infrastructure or taking steps to lower the dwell time at stops.
Transfer waiting times can be reduced by adding frequency or coordinating schedules between tranist routes.

Past studies of transit equity analysis have typically relied on aggregate travel patterns to represent demand, or used accessibility metrics instead of convenience metrics.
These data sources do not provide a complete information of ridership and can ignore certain travel costs such as transfers. Origin-destination data, which is increasingly available to transit agencies, enables a measure of transit equity that is weighted by demand and convenience metrics that include observed travel and wait times.
Furthermore, we use data from on-board rider surveys to examine the differences in trip purpose and modal split between income groups as possible explanatory factors for differences in convenience.
An empirical study using transit data from Boston, Massachusetts is included to demonstrate how these methods can be applied in practice.

\begin{figure}[!ht]
    \centering
    \includegraphics[width=.7\linewidth]{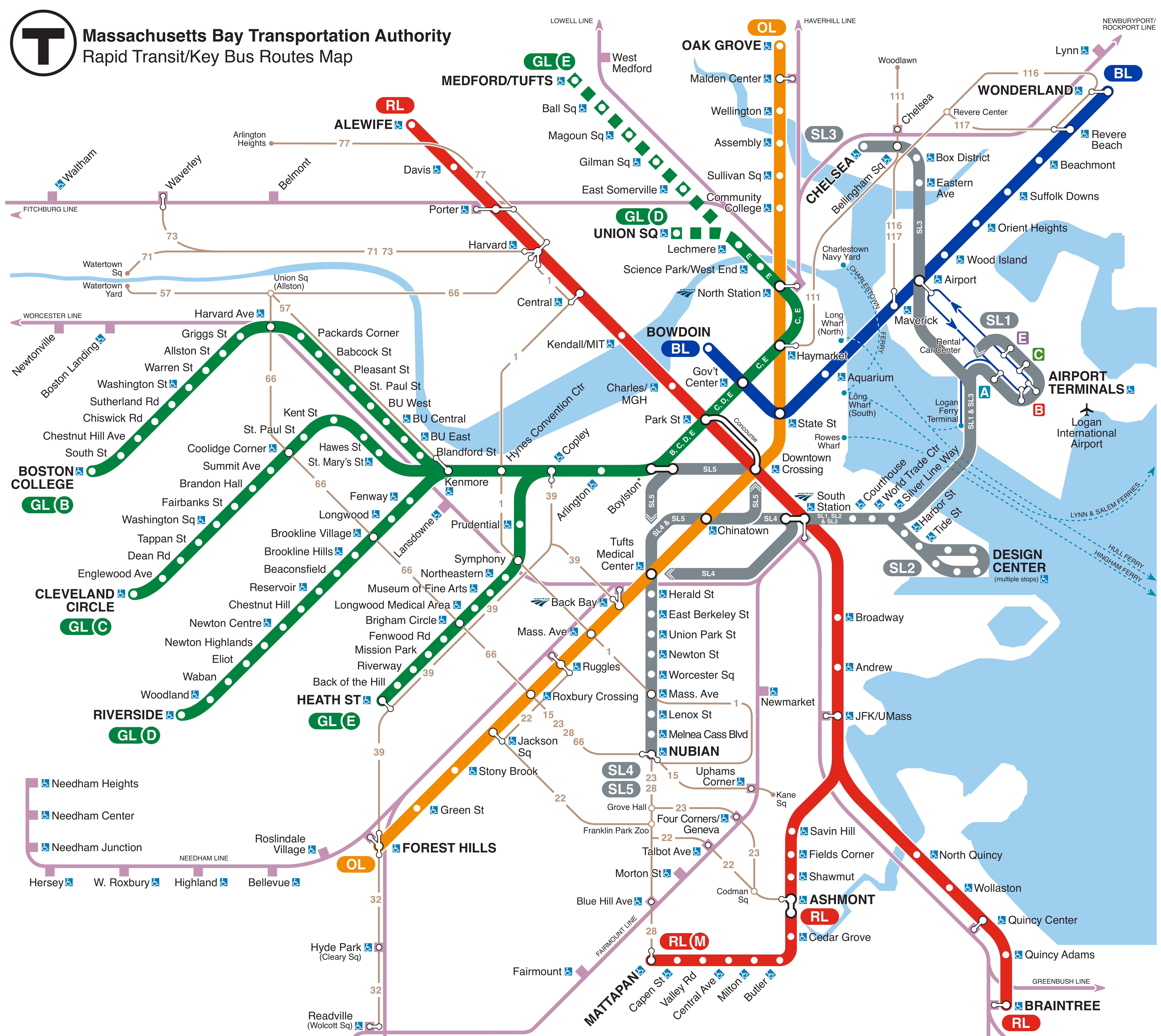}
    \caption{MBTA Rapid Transit and Key Bus Routes Network Map \cite{MBTA_network}.}
    \label{fig:mbta_network}
\end{figure}

Transit service in the Greater Boston area is provided by the Massachusetts Bay Transportation Authority (MBTA).
It is one of the largest transit agencies in the United States, serving 1.4 million trips on an average weekday in September 2019 \cite{MBTA_DASH}. 
It includes four rapid transit lines, over one hundred bus routes, a commuter rail system, a paratransit service and several ferry routes.
The bus and rail network are designed as a hub-and-spoke system, where each line of the rapid transit network converges in the central business district and the bus lines primarily provide access to rail stations. 
The MBTA rapid transit network map is shown in Figure~\ref{fig:mbta_network}
This analysis focuses on the rapid transit and bus network as they collectively serve 90\% of all trips. 

To the best of authors' knowledge, this paper is the first study to propose a disaggregate transit equity analysis that uses journey-level origin-destination data to quantify transit convenience across income groups.
It also provides expansions upon previous empirical studies into transit equity in the Greater Boston area, providing actionable information for local policy makers. 
Implications for the MBTA and transit equity more broadly are discussed in this paper. 

The remainder of the paper is organized as follows. 
The Literature Review section summarizes the state of the art in transit equity research.
The Methodology section describes data sources, the process for synthesizing transit service and demographic information, and the statistical techniques use to quantify transit equities across income groups. 
Empirical study results in the Greater Boston area are presented in the Results section.
Finally, the policy implications, limitations and opportunities for future research are described in the Discussion section.

\section{Literature review}


Transit service equity has been the subject of a wide and very active body of research \citet{litman2002evaluating}.
The literature typically focuses on disparate transit service between low and high income communities, but some studies use additional socioeconomic data to identify ``advantaged'' and ``disadvantaged'' communities. 
\citet{carleton2018comparative} provides an excellent overview of the challenges in defining and measuring equity as it relates to public transit.
There have been many empirical studies that compute equity measures for existing transit networks \cite{griffin2016public, bartzokas2019measuring}, and others methodological studies that use equity as an explicit goal for designing or modifying transit networks \cite{ferguson2012incorporating, gutierrez2017multi}.

Spatial transit equity analyses in the U.S. typically use \emph{transit access} as a metric for convenience. \citet{camporeale2017quantifying} quantified the horizontal and vertical equity in transit planning from a spatial and social perspective, and characterized transit equity as an access to means to any destination from any origin. \citet{aman2020transit}, in a paper introducing a methodology to identify transit deserts, utilized spatial and temporal metrics, describing accessibility by connectivity to the network and connectivity to the destinations. \citet{sharma2020equity} computes transit connectivity and equity using spatial demographic data and transit network characteristics. In its analysis, transit connectivity is defined using ``connective power'' per transit line, which was calculated using speed and route distance from a node to a destination, where the node is defined using a ratio between number of households and employment. \citet{lyons2021transit}, in developing the Transit Economic Equity Index, used the number of stops in a given neighborhood or near an employment center to define transit accessibility. 
Unlike the aforementioned papers that measure the accessibility provided by each line, this paper does not assume any trip purpose or value in expected destinations. Rather, by using origin-destination (OD) data, this paper is able to quantify the transit convenience for observed destination choices in determining how well the transit system is serving low-income communities. 
Research on transit equity using actual OD flows is limited, possibly due to the challenges of collecting such data.
\citet{farber2016space} uses OD data to characterize the spatio-temporal mismatch between transit supply and passenger demand for Greater Salt Lake City, Utah, but does not consider transit convenience with respect to travel speed and transfer.
\citet{lee2019assessing} also uses OD flows and travel times, but is focused on measuring transit competitiveness rather than equity. 
Recently, \citet{dixit2021examining} uses OD patterns along with community demographics to determine whether riders with low-income neighborhoods are served by more circuitous routes than higher income neighborhoods, a related but different measure of convenience than the one presented in this paper.


One important aspect of equity measurement is the determination of disadvantaged communities.
Typical approaches include using national survey data related to income.
For example, \citet{sharma2020equity} defined transit equity using census household income data and the Gini index for cities in Tennessee. 
Transit riders are not necessarily a representative sample of their communities, however. 
This paper uses an on-board survey conducted by the MBTA to identify communities of low-income transit riders, an approach that typically provides a more accurate representation of ridership socio-demographics than census data \cite{karner2015comparison}. 
Other transit equity research has leveraged on-board ridership surveys for locating disadvantaged communities \cite{LUNA, palm2020social}.
More recent sociological research has included mobility patterns to identify so-called ``triple-disadvantaged'' communities \cite{levy2020triple}, a promising approach for future transit equity studies. 

This paper demonstrates how spatial equity can be measured using origin-destination data by presenting a case study for the Greater Boston.
Past papers have also studied the equity of transit service in Boston using different methodologies. 
\citet{LUNA} utilized the MBTA 2015-17 ridership survey to present an equity analysis of the MBTA bus system per route. The transit characteristics examined were on-time performance, dropped trips, and overcrowding. The analysis found that lines with higher rates of low-income riders experienced worse reliability and on-time performance with the bus system. 
\citet{dumas2015analyzing} found that MBTA riders coming from areas that are predominantly Black or African American had longer travel times and lower trip speeds than riders coming from areas that are predominantly White. 
\citet{williams2014measuring} analyzed the transportation equity of the Boston-Cambridge-Newton metropolitan statistical area.
To the authors' knowledge, there is no previous MBTA equity analysis that examines entire passenger journeys to identify socio-economic gaps in transit service. 



\section{Methodology}

In this section, we describe the input data and the statistical methods used for the proposed transit equity analysis.

\subsection{Origin-Destination Data}

Passenger origin-destination data are used in this paper to calculate transit convenience metrics including in-vehicle time, number of transfers, and transfer wait time.
Modern transit agencies are equipped with transit fare collection systems, which provides ''tap-on'' records or both ''tap-on'' and ''tap-off'' records for each passenger journey that interacts with fare boxes. 
For transit agencies without ''tap-off'' data, journey destinations can be inferred for transit systems based on ``tap-on'' information. 
As transit agencies move to provide real-time information to passengers, systems for estimating or collecting origin, destination and transfer information are becoming more common~\cite{figliozzi2018study}.
Note that number of transfers and transfer time are also used for generating convenience metrics used in the proposed transit equity analysis.

The MBTA transit system uses an ``open-loop'' fare system, wherein fare card transactions are registered only when the passenger enters the system.
An origin-destination-transfer inference algorithm, ODX, proposed by \citet{ODX_Gabriel} is implemented within the MBTA to estimate passenger transfers and destinations with a high degree of accuracy.
ODX stands for origin, destination, and transfer inference algorithm, which is an extension of the OD inference algorithm proposed by \citet{ODX_Zhao}.
Data collected with automatic systems including Automatic Vehicle Location (AVL), Automatic Passenger Counting (APC), and Automatic Fare Collection (AFC) data are used as inputs and both destinations and transfers are inferred in the ''tap-on'' only transit system. 
The origin-destination data used in this paper is provided by the MBTA.

In ODX dataset, each OD record represents a single ``ride'', or leg of the journey. 
Each record contains an anonymized passenger identifier, journey identifier, the entrance and exit stop ids, and timestamps.
Multi-leg journeys are linked using the journey identifiers, allowing the number of transfers and transfer waiting time to be computed for each journey.
The OD data used in this study contains records for bus and rail journeys. 
The ODX data used to generate the results presented in this analysis covers January 2019 for which there were over 2,040,000 journey records.

The convenience metrics were then determined for each trip using the timestamps and network information, including number of transfers, in-vehicle travel time, and transfer wait time. 
The number of transfers and in-vehicle travel time are normalized by trip distance in order to avoid the effects of trip length, given that longer trips would be expected to have more transfers and a longer travel time.
The measure of trip length used for normalization is the distance travelled through the transit network, rather than the Euclidean distance between origins and destinations.
The network distance for each origin-destination pair was determined using MBTA's Generalized Transit Feed Specification (GTFS), which provides a geographic representation of the path travelled by transit vehicles. 
The journey records with convenience metrics were then aggregated by origin stops and merge with demographic data described in the following section.

\subsection{Demographic Data}


Unlike most transit equity studies, passenger surveys are used in this paper to estimate the demographics of transit riders.
Compared to census data, survey data provides a direct estimation of the demographics of transit riders, rather than the demographics of the community near each transit stop.
Since 2012, the Federal Transit Administration has required large transit agencies to conduct comprehensive passenger surveys at least once every five years \cite{agrawal2017comparing}.
Agencies are also required to ensure participation from under-represented groups.
Therefore, the proposed transit equity analysis approach in this paper can be applied to any transit agencies with origin-destination data and passenger surveys.

Low-income transit rider communities are identified by the proportion of survey respondents whose reported income is below the defined threshold.
The MBTA defines low-income status as respondents with household incomes less than \$43,500, which is 60\% of the median income for the MBTA service area \cite{survey}. 
Note that the low-income transit rider communities may not be entirely analogous to low-income residential communities.
For example, a transit station near an employment center with low wages might serve primarily low-income riders who are returning home after work, despite the employment center being located in an area with high-income residents. 

The empirical study in this paper uses the MBTA 2015-17 System-wide Passenger Survey \cite{survey} which contains a range of questions including household income, gender, vehicles per capita, trip frequency, trip purpose, minority status, and more. 
The 2015-17 survey is the most recent survey conducted by the MBTA. 
For privacy purposes, the responses are aggregated by station for the rail network, and by route for the bus network, since bus stops have lower ridership.
In order to characterize ridership at the stop level for both bus and rail, the average income statistics of the bus routes were applied to each stop along the routes. 

Overall, 29\% of survey respondents were characterized as low-income. 
The results vary by travel mode; low-income riders represented 42\% of responses from bus riders and 26\% of responses from rail riders.
Approximately 34\% of respondents indicated minority status, and this trend was also more prevalent for bus than for rail. 
Transit riders are more likely to report low-income and minority status than the region as a whole \cite{acs_data}. 
While the survey was administered prior to the study period for the journey-level data, a significant shift in rider demographics between the two periods would not be expected.


\subsection{Data Synthesis}


In order to compare transit service equity across different income groups, it is necessary to merge the convenience metrics with the survey demographics. 
While both sets of input data have a stop-level spatial resolution, larger geographies were used to represent communities, given that people whose trip originates at a given stop also has access to nearby transit service.
The MBTA service area was therefore divided into small geographic areas of approximately equal population and both the transit convenience and ridership demographics were then assigned to those areas using a buffer method described below.
Census blocks and block groups were used to define the geographic areas for convenience.
The impact of aggregating the transit and ridership data at the census block or block group level is also discussed in the results section. 

The stop-level convenience metrics and demographics were aggregated to census blocks using a buffer method: any stops that could be accessed from each block were included in the calculation of that block's characteristics.
A 500 foot buffer was drawn around the boundary of each census block, representing a reasonable walking distance to access transit \cite{untermann1984accommodating}.
Furthermore, different buffer sizes are tested in the results section.
The convenience metrics and demographics of each census block was generated by taking the weighted average based on stop-level ridership over every stop within the block boundaries and buffer areas.
The convenience metrics and demographics of census block groups are aggregated following the same steps as census blocks.
The synthesized data yields approximately 4,900 census block groups and 39,000 census blocks that are covered by MBTA services within the Greater Boston area. 

\subsection{Statistical Methods}


A linear regression approach was utilized in this paper to test the hypothesis that low-income riders experience worse travel experience within the transit system, indicating by transit convenience measures including in-vehicle time per mile, transfers per mile and transfer wait time per mile. 
For each census block or block group, the proportion of low income riders is served as the dependent variable and the convenience measures are the explanatory variables in the regression model.
Two additional travel characteristics are included in the analysis: network distance and share of travel distance spent on rail.
These characteristics are introduced in the analysis for the following reason: the convenience differences between low-income and high-income riders could be explained by travel distances and mode share of the rail system; we have hypotheses that low-income riders have longer trips and use less rail services compared to high-income riders, which lead to worse convenience metrics.
The parameters in regression models are estimated using the Ordinary Least Squares (OLS) method. 

\section{Empirical results}


In this section, we present the results from regression models which tests the relationship between low-income ridership and transit convenience. We also present a spatial unit comparison analysis to demonstrate the variation of results over a larger geographical unit. Finally, we present a sensitivity analysis over the buffer size chosen in the proposed approach. . 

\subsection{Block Level Analysis}

First, we show the transit equity analysis results given census blocks as spatial unit of aggregation with data from January 2019. 
There exists a positive and statistically significant relationship between low-income ridership and the three convenience metrics, travel time per mile, transfers per mile, and transfer waiting time, based on our regression models presented in Table~\ref{tab:table1}. 

\begin{table}[h!]
  \begin{center}
    \caption{Results of regression models predicting convenience outcomes from low-income ridership by Census Block in January 2019.}
    \label{tab:table1}
    \begin{tabular}{|l|c|c|c|}
    \hline
      \textbf{Explanatory Variables} & \textbf{Parameter} & \textbf{t-value} & \textbf{p-value} \\
      \hline
      Time by Distance (min/mile) & 0.6791 & 10.751 & < 2.2e-16***\\
      Transfers by Distance (\#/mile) & 0.0378 & 11.045 & < 2.2e-16***\\
      Transfer Wait Time (min) & 0.0301 & 4.9785 & 6.604e-07***\\
      Distance (mile) & -0.0168 & -2.4490 & 0.0152*\\
      Rail Mode Share (\%) & -0.2000 & -6.8428 & $<$ 8.785e-12***\\
      \hline
       \textbf{Adjusted R-squared:}   & 0.0243 & \textbf{R-squared:}  & 0.0246 \\
      \hline
    \end{tabular}
    \item Statistical significance coded as *p $<0.05$ , **p $<0.01$ , ***p $<0.001$ 
  \end{center}
\end{table}

The transfers by distance parameter, measured in number of transfers per mile, demonstrates a strong correlation with low-income ridership. 
For every 1\% increase in low-income ridership within a census block, there is a .04 transfer increase per mile for trips originating from the census block.
This suggests that riders whose trips begin in locations with more low-income riders have to transfer more often in order to travel the same distance. 

The in-vehicle time by distance parameter, measured in minutes per mile, also demonstrates a strong correlation with low-income ridership. 
For every 1\% increase in low-income ridership within a census block, there is, approximately, a 0.68 minute (41 second) increase in in-vehicle time per mile originating from that census block.
This suggests that lower income travel must spend more time to travel a given distance compared to riders from higher income communities. 

To further illustrate the difference at the community level between higher income and lower income areas, blocks with less than 25\% low-income ridership were designated as high-income and blocks with more than 50\% low-income ridership were designated as low-income. The distribution of income by census tracts can be seen in Figure~\ref{fig:low_income} below. We visualized lower income communities and convenience metrics at census tract level for the clarity purpose given there are 39,000 census blocks within the Greater Boston area. 

\begin{figure}[!ht]
    \centering
    \includegraphics[width=.7\linewidth]{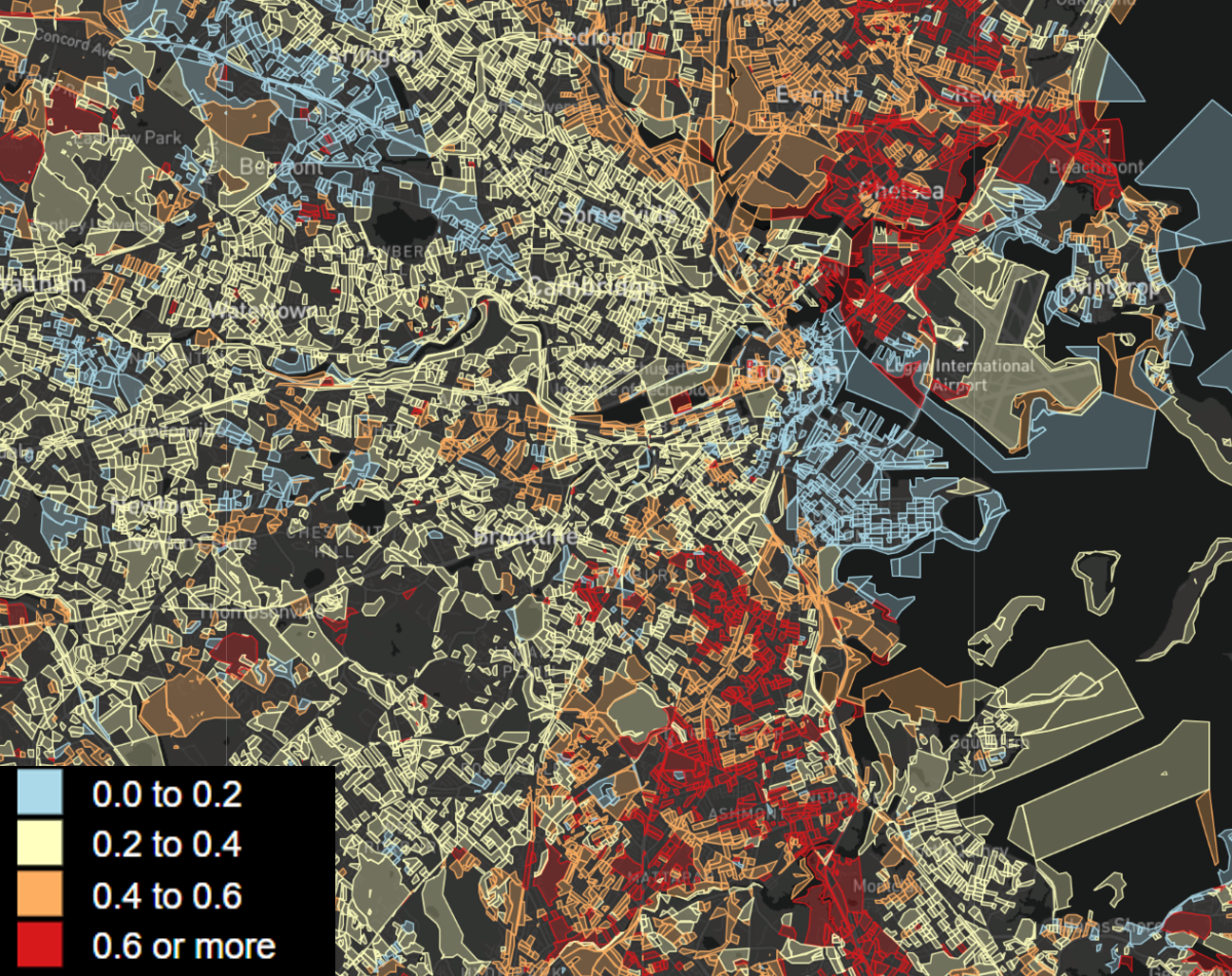}
    \caption{Distribution of low income transit rider communities in Greater Boston. Numbers in the legend represent the proportion of low-income ridership in census tracts. }
    \label{fig:low_income}
\end{figure}

\begin{figure*}[t!]
    \centering
    \begin{subfigure}[b]{0.49\textwidth}
        \centering
        \includegraphics[width=.96\linewidth]{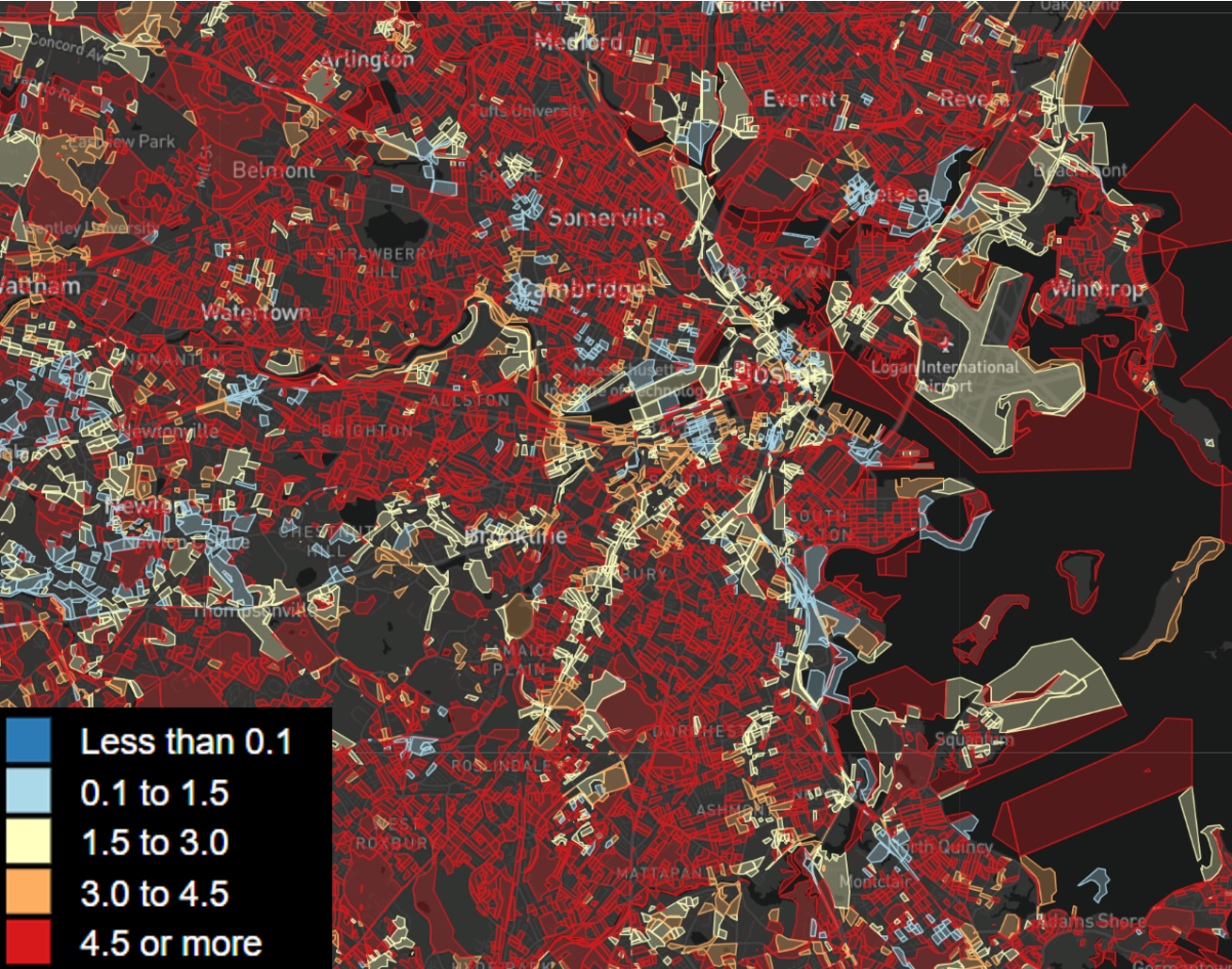}
        \caption{Number of transfers per mile}
        \label{fig:transfers}
    \end{subfigure}%
    \begin{subfigure}[b]{0.49\textwidth}
        \centering
        \includegraphics[width=.95\linewidth]{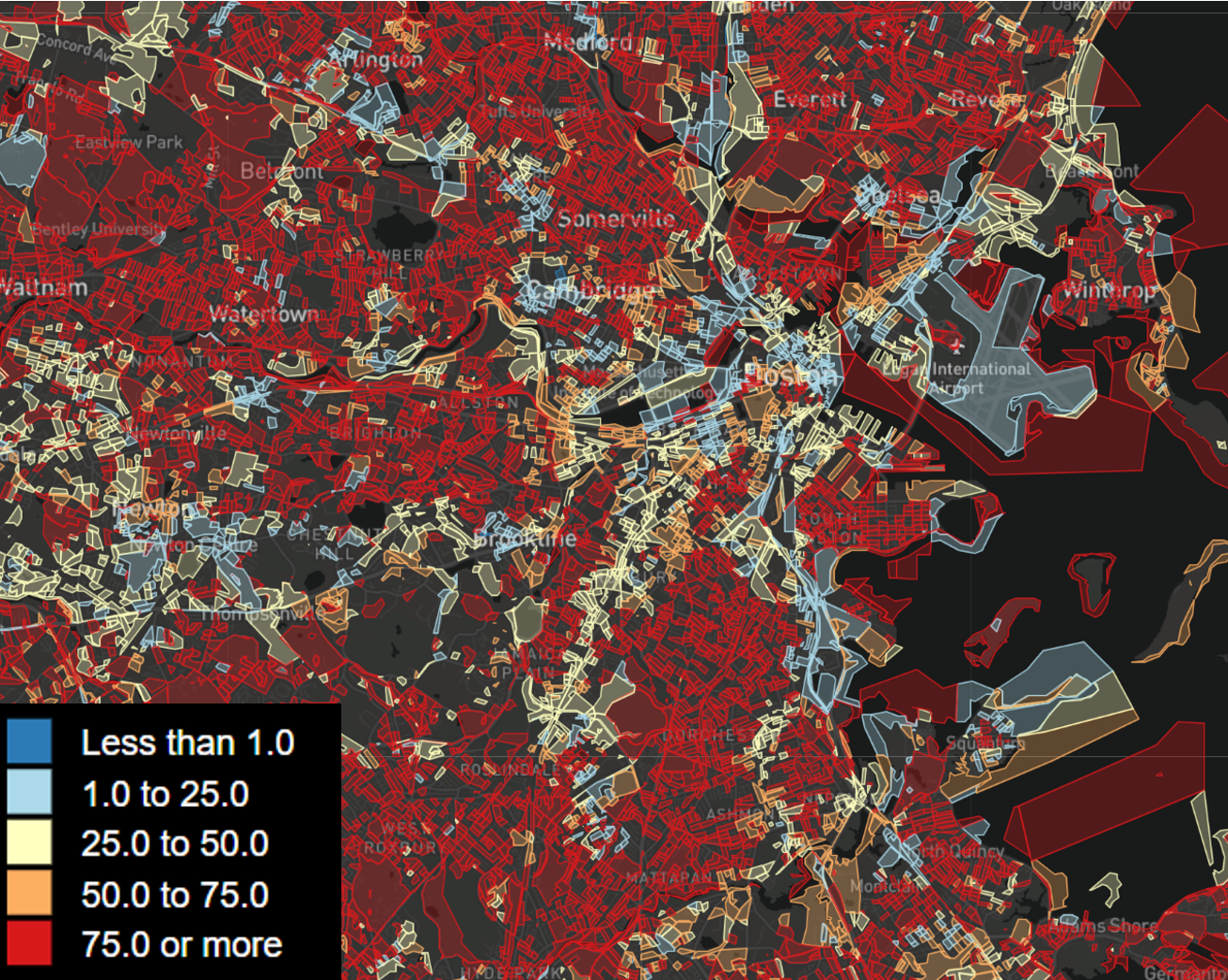}
        \caption{Travel time per mile}
        \label{fig:travel_time}
    \end{subfigure}
    \caption{Census tract level convenience metrics.}
    \label{fig:census_block_analyses}
\end{figure*}

Then, the different convenience metrics were plotted for each census tracts, as shown in Figure~\ref{fig:transfers}. 
It can be observed that the low-income communities to the south of downtown Boston also have a high number of transfers per mile, while wealthier communities to the west have generally fewer transfers. 
Low-income census blocks had about 6.9 transfers per mile where high-income census blocks had about 4.0 transfers per mile.
The results for travel time normalized by trip distance are shown in Figure~\ref{fig:travel_time}.
Low-income census blocks had an average of 84 minutes per mile where high-income census blocks had about half the time, with an average of 43 minutes per mile. 

A potential explanation for the longer travel time is the longer transfer wait time. With each 1\% increase in low-income ridership through a census block, there is approximately a .03 minute, or a 2 second increase in the wait time. Although this is a smaller relational estimate, the relationship is statistically significant. Thus, we can assume that the transfer wait times contribute to the longer travel times. 

\subsection{Block Group Level Analysis}
Given a census block group is a larger spatial unit, as a collection of census blocks, we would like to know whether the patterns for census blocks held for larger communities. A census block group typically has around 600 to 3000 inhabitants, and between 250 and 550 housing units. The results demonstrated below is the same analysis as the census block level, except each spatial unit is a block group.

\begin{table}[h!]
  \begin{center}
    \caption{Results of regression models predicting convenience outcomes from low-income ridership by Census Block Group in January 2019.}
    \label{tab:table2}
     \begin{tabular}{|l|c|c|c|}
     \hline
      \textbf{Explanatory Variables} & \textbf{Parameter} & \textbf{t-value} & \textbf{p-value} \\
      \hline
      Time by Distance (min/mile) & 0.8740 & 6.8886 & 1.594e-11***\\
       Transfers by Distance (\#/mile) & 0.0598 & 6.2850 & 6.822e-10***\\
      Transfer Wait Time (min) & 0.0026 & 0.2063 & 0.8366\\
      Distance (mile) & -0.0087 & -2.5105 & 0.0124*\\
      Rail Mode Share (\%) & -0.0076 & -5.594 & 3.551e-08***\\
      \hline
       \textbf{Adjusted R-squared:}   & 0.07047 & \textbf{R-squared:}  & 0.06892 \\
      \hline
    \end{tabular}
    \item Statistical significance coded as *p $<0.05$ , **p $<0.01$ , ***p $<0.001$ 
  \end{center}
\end{table}

The results above are relatively similar to the census block analysis. The models for time by distance and transfers by distance have approximately a .2 min/mile difference, both with p-values less than $0.001$. The transfer wait time is not statistically significant in the block group level analysis. The distance models have approximately a 0.01 mile difference, and are both negatively correlated with low-income ridership, with a significance of p less than 0.05. Finally, rail mode share is significantly correlated with low-income ridership share both at the block level and at the block group level.

\subsection{Sensitivity Analysis}
The above analysis utilized a 500ft buffer when aggregating stop-level information into census blocks and block groups. 
If a census block or block group falls within a stop buffer, the stop's origin-destination and demographics data is included in that block or block group. In this subsection, we ran the same analysis for census block groups with a 1000ft buffer to test the impact of buffer sizes.

\begin{table}[h!]
  \begin{center}
    \caption{Results of regression models predicting convenience outcomes from low-income ridership by Census Block Group with a 1000ft buffer in January 2019. }
    \label{tab:table4}
     \begin{tabular}{|l|c|c|c|}
    \hline
      \textbf{Explanatory Variables} & \textbf{Parameter} & \textbf{t-value} & \textbf{p-value} \\
      \hline
      Time by Distance (min/mile) & 0.0105 & 4.9877 & 1.225e-06***\\
      Transfers by Distance (count/mile) & 0.0574 & 6.2088 & 1.002e-09***\\
      Transfer Wait Time (min) & -0.0081 & -0.6735 & 0.5007\\
      Distance (mile) & -0.0093 & -2.892 & 0.0040**\\
      Rail Mode Share (\%) & -0.7758 & -5.818 & 9.992e-09 ***\\
       \hline
       \textbf{Adjusted R-squared:}   & 0.06547 & \textbf{R-squared:}  & 0.0639 \\
      \hline
     \end{tabular}
    \item Statistical significance coded as *p $<0.05$ , **p $<0.01$ , ***p $<0.001$ 
  \end{center}
\end{table}

The 1000ft buffers for census block groups include more ridership compared to 500ft buffers. The time by distance, transfers by distance, distance, and rail mode share estimates are in the same direction as the 500ft buffer estimates, with significance. Furthermore, the wait time under both the 1000ft buffer model and the 500ft buffer model, are not significant according to p-values. The coefficients in both regression models are similar: the transfers by distance is around 0.06 in the 500ft model and also 0.06 in the 1000ft model. 
This analysis results demonstrate that the size of the buffer does not have a significant impact over regression models.

\subsection{Trip Purpose}
One possible explanation for the differences in transit convenience we are seeing between low-income riders and high-income riders could be the vast differences in trip purpose. 

\begin{table}[h!]
  \begin{center}
    \caption{Results of regression models predicting trip purpose from low-income ridership}
    \label{tab:table1}
    \begin{tabular}{|l|c|c|c|}
    \hline
      \textbf{Explanatory Variables} & \textbf{Parameter} & \textbf{t-value} & \textbf{p-value} \\
      \hline
      Home to Work / Work to Home & -0.4751 & -13.959 & < 2.2e-16***\\
      \hline
      Home to Other / Other to Home & 0.2367 & 12.220 & < 2.2e-16***\\
      \hline
      Other (not a home trip) & 0.0575 & 4.2394 & 4.062e-05***\\
      \hline
      Home to Social / Social to Home & 0.0317 & 2.6496 & 0.0090**\\
      \hline
      Home to School / School to Home & 0.1496 & 8.9240 & 2.357e-15***\\
      \hline
       \textbf{Adjusted R-squared:} & 0.5144 & \textbf{R-squared:}  & 0.5179 \\
      \hline
    \end{tabular}
    \item Statistical significance coded as *p $<0.05$ , **p $<0.01$ , ***p $<0.001$ 
  \end{center}
\end{table}

Low-income riders are more likely than non-low-income riders to use public transit to travel on non-home trips and to travel outside of work. There is also a significant negative relationship between low-income riders and home/work trips. This suggests further that low-income riders are using public transit for non-commute trips more often than for commute trips. 

\subsection{COVID Analysis}
With the introduction of the COVID pandemic, the MBTA cut their services. We expect this to affect ridership patterns and equity of service. 

\begin{table}[h!]
  \begin{center}
    \caption{Results of regression models predicting convenience outcomes from low-income ridership by Census Block with a 500ft buffer in January 2020 }
    \label{tab:table4}
     \begin{tabular}{|l|c|c|c|}
    \hline
      \textbf{Explanatory Variables} & \textbf{Parameter} & \textbf{t-value} & \textbf{p-value} \\
      \hline
      Time by Distance (min/mile) & 0.7210 & 12.66 & < 2.2e-16 ***\\
      Transfers by Distance (count/mile) & 0.0243 & 9.023 & < 2.2e-16***\\
      Transfer Wait Time (min) & -0.0319 & -5.637 & 1.808e-08***\\
      Distance (mile) & -0.0079 & -6.438 & 1.346e-10 ***\\
       \hline
       \textbf{Adjusted R-squared:}   & 0.06547 & \textbf{R-squared:}  & 0.0639 \\
      \hline
     \end{tabular}
    \item Statistical significance coded as *p $<0.05$ , **p $<0.01$ , ***p $<0.001$ 
  \end{center}
\end{table}
January 2020 and January 2019 both showed a negative correlation with low-income communities and distance travelled. 
January 2020 had a 0.05 larger estimate of minutes per mile, suggesting the typical trip for low-income communities was 0.05 minutes longer in 2020 than in 2019. 
However, January 2020 had a decrease in the number of transfers by .005. 
These results suggest there was a large similarity in trip convenience in January 2020 and January 2019. 

\begin{table}[h!]
  \begin{center}
    \caption{Results of regression models predicting convenience outcomes from low-income ridership by Census Block with a 500ft buffer in October 2020 }
    \label{tab:table4}
     \begin{tabular}{|l|c|c|c|}
    \hline
      \textbf{Explanatory Variables} & \textbf{Parameter} & \textbf{t-value} & \textbf{p-value} \\
      \hline
      Time by Distance (min/mile) & 0.8033 & 12.66 & 0.0003***\\
      Transfers by Distance (\#/mile) & 0.0341 & 9.023 & 9.86e-06***\\
      Transfer Wait Time (min) & 0.0240 & 4.5705 & 4.954e-06***\\
      Distance (mile) & -0.0168 & -6.438 & 0.01523*\\
       \hline
       \textbf{Adjusted R-squared:}   & 0.06547 & \textbf{R-squared:}  & 0.0639 \\
      \hline
     \end{tabular}
    \item Statistical significance coded as *p $<0.05$ , **p $<0.01$ , ***p $<0.001$ 
  \end{center}
\end{table}

The estimated wait time increased by 0.05 minutes (or 3 seconds) between January 2020 and October 2020, with a p-value of < .001. This suggests that communities with a higher number of low-income riders experienced an increase in transfer wait time over the course of the pandemic. 

To illustrate this difference more concretely, we compared two census tracts with varying levels of low-income riders in January 2019. Roslindale is a lower income town within the Boston metropolitan region and Belmont is a higher income town. In census tract 25025110502 in the center of Roslindale, 29.9\% of riders are low-income. In census tract 25017357400 in the center of Belmont, 20.6\% of riders are low-income. An average distance a rider travels coming from Roslindale is .22 miles, where an average distance traveled coming from Belmont is .27 miles. The average number of transfers in Roslindale was ~1.1 transfers higher than in Belmont. The time spent in transit was on average 5\% higher in Roslindale than in Belmont. 

After the service cuts coming from COVID, in October 2020, we found a dramatic increase in the difference of time spent per journey. The average distance originating from Roslindale changed by less than .04 miles, at an average of .25 miles. The average distance originating from Belmont remained the same at .28 miles. The average number of transfers is similar, and in fact the difference was reduced during covid. Trips originating out of Roslindale had only a .4 transfer difference than trips originating from Belmont, in comparison to the 1.1 transfer difference in January 2019. However, the average time spent in transit coming from Roslindale is 26\% higher vs. the 5\% pre-covid. With a difference of ~ 9\% low-income ridership, there is a dramatic time difference in transit. This suggests the MBTA's COVID cuts disproportionally affected riders coming from lower-income communities.

\section{Discussion}

This paper presents a methodology for measuring the transit equity between income groups using two data sources that have not yet been combined in transit equity research: on-board ridership surveys and passenger origin-destination flows.
Our proposed approach provides a better estimation of ridership demographics while also permitting new measures of transit service convenience: number of transfers per unit distance, transfer wait time and travel time per unit distance.
A case study in the Greater Boston area using MBTA ridership data is conducted in this paper. A regression model is established to show that the proportion of low-income transit riders in a community (census blocks or block groups) has a significant positive correlation with transit inconvenience in the Greater Boston area. 

This research has implications for transit agencies, and specifically for transit service in the Greater Boston area.
Measuring convenience and equity using passenger journey data allows transit agencies to identify opportunities to improve service for low-income riders through service changes.
The additional transfers per mile faced by low-income riders can be addressed by designing more direct routes.
Longer transfer wait times can be mitigated by adding timed transfers to the schedule at locations where low-income riders are more likely to transfer. 
Travel times per unit distance are also subject to improvement through adding targeted transit priority infrastructure such as exclusive rights-of-way, queue jump lanes and signal priority.

It is worth mentioning that this paper is not intended as a critique of transit service and planning in Greater Boston. 
Low and high-income areas are not fixed over time, and their distributions often change in response to new or improved transit service \cite{perk2013silver, cao2018and}, making it difficult for transit agencies with limited resources to maintain perfect equity throughout the system.
This paper intends to provide a journey-based approach that can be used by agencies to evaluate existing service, track progress and plan future changes.

There are some limitations to the methods and data presented in this paper.
The MBTA 2015-17 on-board ridership survey aggregates responses for each bus route in the network in order to preserve the privacy of the respondents.
The aggregation of bus route data along bus routes led to a distribution assumption that all stops along a line had similar characteristics. 
This could have contributed to skewed results for the community income distribution. 
Note that responses for rail trips are aggregated by stop and therefore do not suffer from this limitation.
Furthermore, 99\% of the surveys returned were filled out using the English version of the form, which implies undercounting of minority communities that are non-English speaking. 
Finally, a fixed-distance buffer method was used to identify the transit stops that could be accessed from each census block or block group.
This simplification ignores the heterogeneity in the distance that people are comfortable traveling in order to access transit stops. 

There are many promising opportunities for future research based on this study.
First, measures of income are important for identifying disadvantaged communities, but other socioeconomic or mobility data could be used for a more comprehensive analysis.
Furthermore, additional measures of convenience could also be included in an equity analysis, such as initial wait time.
Finally, comparing the results before, during and after the COVID-19 pandemic, which had a tremendous impact on public transit demand and service, could provide additional insights into how pandemic-related service changes affected low-income riders and how equity can be improved during the post-COVID recovery period.

\section{Acknowledgements}
The authors would like to thank Anson Stewart, Anna Gartsman, Alissa Zimmer and Professor Jinhua Zhao for their comments on the early stages of this research. 
The authors would also like to thank the MBTA for their generous support of this project.

\section{Author Contribution Statement}
The authors confirm contribution to the paper as follows: study conception and design: D.S., X.G., N.S.C.; data collection: D.S., X.G., N.S.C.; analysis and interpretation of results: D.S., X.G., N.S.C.; draft manuscript preparation: D.S., X.G., N.S.C. All authors reviewed the results and approved the final version of the manuscript.  
The authors do not have any conflicts of interest to declare.

\newpage
\bibliographystyle{trb}
\bibliography{references}
\end{document}